\documentclass[twocolumn,amsmath,amssymb,aps,longbibliography,floatfix]{revtex4-1}
\usepackage{cancel}
\usepackage{enumitem}
\usepackage[utf8x]{inputenc}
\usepackage{graphicx}% Include figure files
\usepackage{dcolumn}% Align table columns on decimal point
\usepackage{bm}% bold math
\usepackage{hyperref}% add hypertext capabilities
\usepackage[switch]{lineno}% Enable numbering of text and display math
\usepackage{float}             % Allows for floating table and figures
\usepackage{subfigure}
\usepackage{tikz}

\begin{document}

\title{The limit of spin lifetime in solid-state electronic spins}

\author{$^{1}$Alessandro Lunghi}
\email{lunghia@tcd.ie}
\author{$^{1}$Stefano Sanvito}

\affiliation{$^{1}$School of Physics, AMBER and CRANN, Trinity College, Dublin 2, Ireland}

\begin{abstract}
{\noindent \bf The development of spin qubits for quantum technologies requires their protection from the main source of finite-temperature decoherence: atomic vibrations. Here we eliminate one of the main barriers to the progress in this field by providing a complete first-principles picture of spin relaxation that includes up to two-phonon processes. Our method is based on machine learning and electronic structure theory and makes the prediction of spin lifetime in realistic systems feasible. We study a prototypical vanadium-based molecular qubit and reveal that the spin lifetime at high temperature is limited by Raman processes due to a small number of THz intra-molecular vibrations. These findings effectively change the conventional understanding of spin relaxation in this class of materials and open new avenues for the rational design of long-living spin systems.}
\end{abstract}

\maketitle

The observation of matter in a coherent superposition of quantum states is a core fingerprint of quantum mechanics. While the laws of Physics allow any object, regardless of its size, to be prepared in such state, the unwanted interaction of a quantum system with a large number of other degrees of freedom causes the destruction of coherence and restore the classical picture of the world \cite{Hornberger2004}. Spins, either nuclear or electronic, are naturally loosely coupled to other degrees of freedom. The coherent control of their quantum states has been achieved in several physical systems, including both solid-state semiconductors \cite{Awschalom2013} and molecules \cite{Wernsdorfer2019}. Several strategies designed to preserve the coherence time, $T_{2}$, from the effect of spurious magnetic interactions have been successful implemented
\cite{Zadrozny2015,Shiddiq2016}. However, spins also inevitably interact with the atomic motion and relax to a non-coherent thermal equilibrium state on a time scale, $\tau$, also called $T_{1}$ or spin-lattice relaxation time. This is the ultimate limit for the coherence time. 

Here we provide a first-principles description of these relaxation processes and show the limits that solid-state vibrations pose to the spin lifetime. Despite the central importance of spin-lattice relaxation for a broad range of disciplines, such as magnetism \cite{Maehrlein2018}, quantum computation \cite{Astner2018} and magnetic resonance \cite{Caravan2006}, its parameter-free first-principles theoretical description is almost unexplored due to the associated daunting computational requirements\cite{Escalera-Moreno2017,Goodwin2017,Astner2018,Lunghi2019b}. In this work we solve the challenge by designing a machine-learning-accelerated \cite{Lunghi2019,Lunghi2020} first-principles strategy able to account for both one- and two-phonon spin relaxation processes. Multi-phonon contributions have so far been beyond the reach of computational methods and their inclusion provides a complete picture of spin relaxation \cite{Lunghi2019b}. We anticipate that these results will enable new routes for the design of long-living spin systems.

We will use as a test case the study of the VO(acac)$_{2}$ molecular qubit embedded in a solid-state crystal \cite{Tesi2016}. This V$^{4+}$-based molecule bears both an electronic, $\vec{\mathbf{S}}$, and a nuclear spin, $\vec{\mathbf{I}}$, and represents a typical building block for quantum computing platforms \cite{Childress2006,Muhonen2014,Godfrin2017a}.

\begin{figure}
\begin{center}
\includegraphics[scale=1]{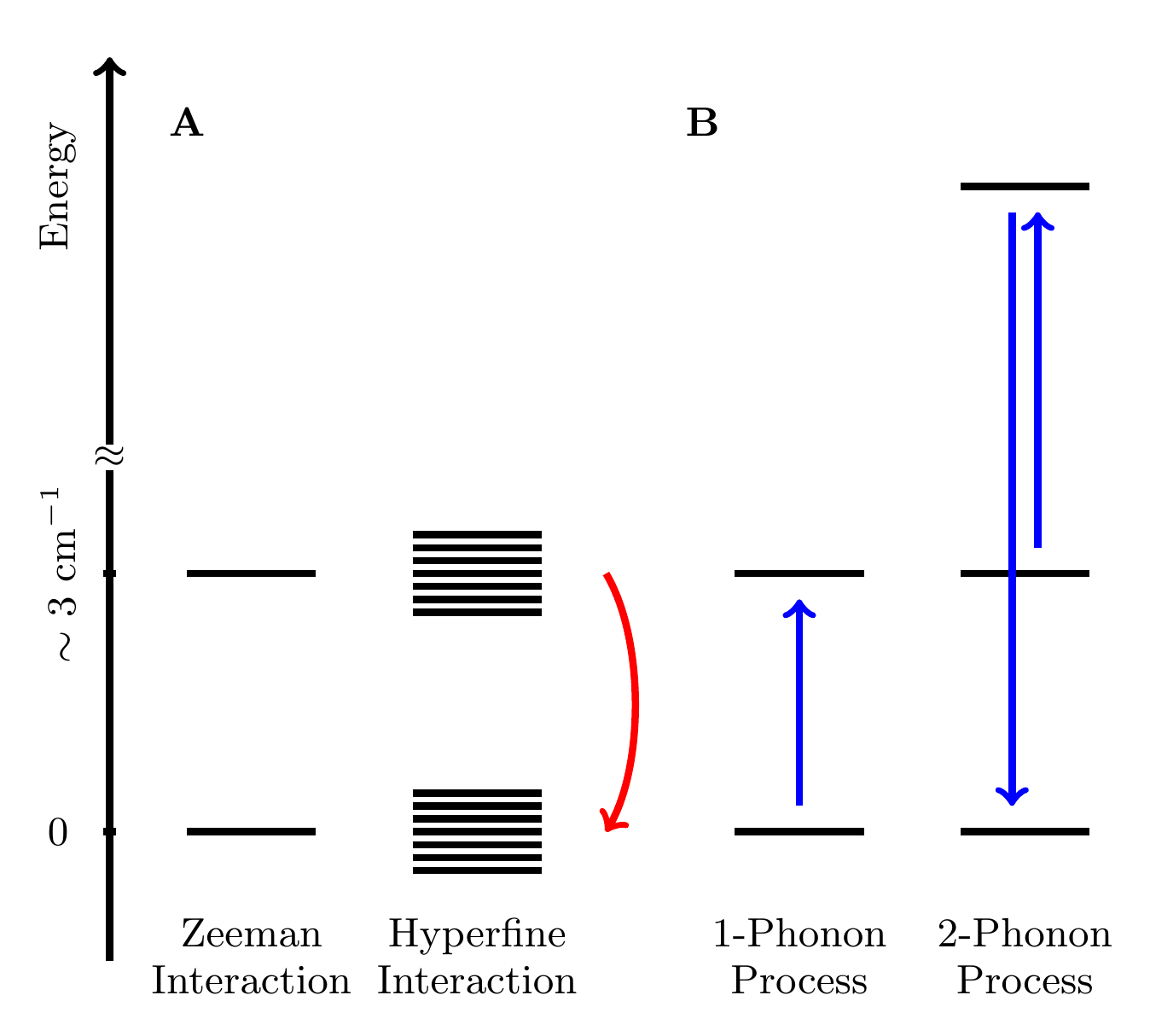}
\end{center}
\caption{\textbf{Schematic Representation of Spin Energy Levels and Spin-Phonon Coupling Processes.} 
(\textbf{A}) The effect of the Zeeman and hyperfine interaction on the spin energy levels with an external 
field of $\sim$ 3 T. (\textbf{B}) An electronic spin-flip from an excited state to the ground state can be 
accompanied by the emission of a phonon with energy in resonance with the spin transition (one-phonon 
process) or the simultaneous absorption and emission of two high-energy phonons, whose energy difference 
is in resonance with the spin transition (two-phonon process).}
\label{scheme}
\end{figure}

In the absence of others spins from the ones explicitly considered, a situation also known as magnetic diluted 
condition, the spin dynamics of a VO(acac)$_{2}$ unit is driven by the spin Hamiltonian, 
\begin{equation}
\hat{H}_\mathrm{s}=\beta_{e}\vec{\mathbf{B}}\cdot\mathbf{g_{e}}\cdot\vec{\mathbf{S}}+\beta_{n}\vec{\mathbf{B}}\cdot\mathbf{g_{n}}\cdot\vec{\mathbf{I}}+\vec{\mathbf{S}}\cdot\mathbf{A}\cdot\vec{\mathbf{I}}\:,
\label{SH}
\end{equation}
where the electronic (S=1/2) and nuclear (I=7/2) spins of $^{51}$V interact with each other through the 
hyperfine tensor, $\mathbf{A}$, and with the external magnetic field, $\vec{\mathbf{B}}$, through the 
electronic (nuclear) gyromagnetic tensor, $\beta_{e}\mathbf{g_{e}}$ ($\beta_{n}\mathbf{g_{n}}$). As 
schematically shown in Fig.~\ref{scheme}A, the Zeeman and hyperfine interactions split the 
spin spectrum, $\omega$, in the $10^{-2}-10$~cm$^{-1}$ range, depending on the size of the external 
field. For energies comparable to $k_\mathrm{B}T$, where $k_\mathrm{B}$ is the Boltzman constant, 
only these states will be populated, as the first excited electronic state is more than 10,000 cm$^{-1}$ 
higher in energy.

In the absence of external perturbations, the state of $\vec{\mathbf{S}}$ and $\vec{\mathbf{I}}$, described 
by the spin-density matrix, $\hat{\rho}$, would evolve coherently in time according to the Liouville equation,
$i\hbar(d\hat{\rho}/dt)=[\hat{H}_\mathrm{s},\hat{\rho}]$. However, in a typical solid-state environment, such as the 
VO(acac)$_{2}$ molecular crystal, the spin lifetime has a finite value due to the coupling of spins with phonons 
$\hat{Q}_{\alpha \mathbf{q}}$, namely with the oscillations of the atoms' positions with frequency,
 $\omega_{\alpha\mathbf{q}}$, and reciprocal lattice vector $\mathbf{q}$. According 
to Redfield theory \cite{Redfield1965}, the spin-relaxation rate, $\tau^{-1}$, due to the cooperative effect of $n$ phonons 
depends on the spin-phonon coupling coefficients, $V_{n-\mathrm{ph}}$, and the Fourier transform of the $n$-phonon 
correlation function, $G^{n-\mathrm{ph}}$,
\begin{equation}
\tau^{-1}\propto V_{n-\mathrm{ph}}^{2} G^{n-\mathrm{ph}} = \left(\frac{\partial^{n} \mathbf{V}}{\partial Q_{1}...\partial Q_{n}} \right)^{2} G^{n-\mathrm{ph}} \:,
\label{rate}
\end{equation}
where $\mathbf{V}$ can be any of the tensors entering in the definition of $\hat{H}_\mathrm{s}$ and $G^{n-\mathrm{ph}}$ 
contains the phonons thermal population, $\bar{n}_{\alpha\mathbf{q}}$, and a condition imposing energy conservation.

Here we consider one- and two-phonon processes at the first-order of perturbation theory. One-phonon processes account for direct spin relaxation and contribute to the relaxation rate through $V_{1-\mathrm{ph}}=(\partial \mathbf{A} / \partial Q_{\alpha\mathbf{q}})$ or $V_{1-\mathrm{ph}}=(\partial \mathbf{g_{e}} / \partial Q_{\alpha\mathbf{q}})$, and $G^{1-\mathrm{ph}}(\omega,\omega_{\alpha\mathbf{q}})\propto\delta(\omega-\omega_{\alpha\mathbf{q}})\bar{n}_{\alpha\mathbf{q}}+\delta(\omega+\omega_{\alpha\mathbf{q}})(\bar{n}_{\alpha\mathbf{q}}+1)$. The first (second) term of $G^{1-\mathrm{ph}}$ describes spin transitions involving the absorption (emission) of a 
single phonon from (to) the thermal bath. Figure~\ref{scheme}B schematically describe the emission process. 
Two-phonon contributions, instead, involve the simultaneous modulation of the spin Hamiltonian by two phonons
and open up Raman relaxation pathways. The relaxation rate due to simultaneous absorption/emission of two phonons, also depicted in Fig.~\ref{scheme}B, depends on 
second-order spin-phonon coupling coefficients, \textit{i.e.} $V_{2-\mathrm{ph}}=(\partial^{2} \mathbf{A} / \partial Q_{\alpha\mathbf{q}}\partial Q_{\beta\mathbf{q'}})$ or $V_{2-\mathrm{ph}}=(\partial^{2} \mathbf{g_{e}} / \partial Q_{\alpha\mathbf{q}}\partial Q_{\beta\mathbf{q'}})$, and $G^{2-\mathrm{ph}}(\omega,\omega_{\alpha\mathbf{q}},\omega_{\beta\mathbf{q'}}) \propto \delta(\omega-\omega_{\alpha\mathbf{q}}+\omega_{\beta\mathbf{q'}}) \bar{n}_{\alpha\mathbf{q}}(\bar{n}_{\beta\mathbf{q'}}+1)$.

\begin{figure*}
\begin{center}
 \includegraphics[scale=1]{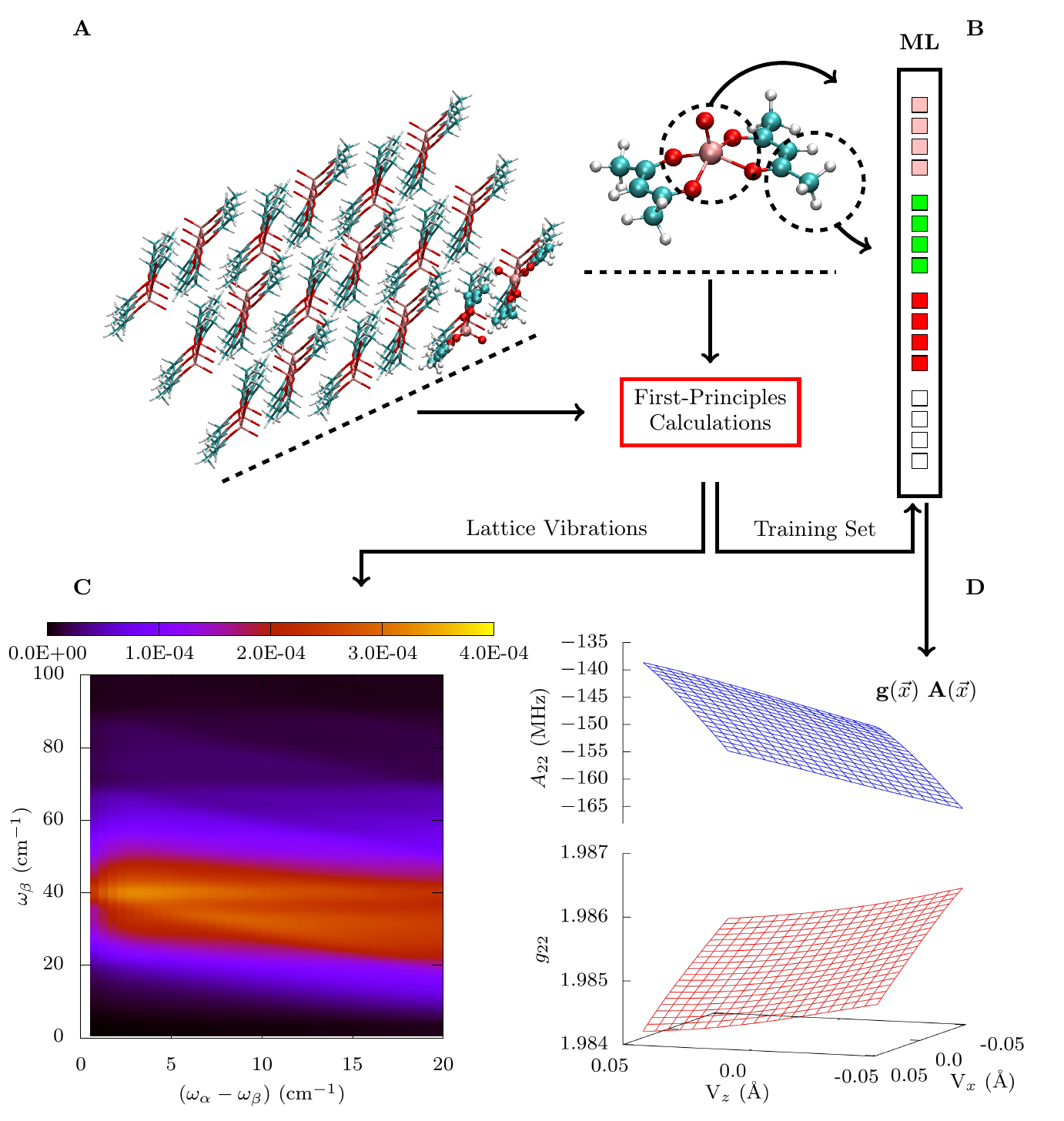}
\end{center}
\caption{\textbf{First-Principles and ML Approach to Lattice and Spin Dynamics.} (\textbf{A}) The 
3$\times$3$\times$3 replica of the VO(acac)$_{2}$ primitive cell used for the simulation of the 
crystal's vibrational properties and the structure of the isolated molecular unit used to generate 
the training set for the ML algorithm. (\textbf{B}) The schematic structure of the ML algorithm 
used to predict the magnetic properties as a function of the general atomic displacements. Each 
atomic environment is converted into a vector of fingerprints that determine the atomic contributions 
to the \textbf{A} and $\mathbf{g_{e}}$ tensors. (\textbf{C}) The Fourier transform of the two-phonon 
correlation function, $G^{2-\mathrm{ph}}$, integrated over the Brillouin zone. (\textbf{D}) Examples 
of ML predictions for the hyperfine and Land\`e tensors as function of the V atomic displacements 
along $x$ and $z$.}
\label{scheme2}
\end{figure*}

\begin{figure*}
\begin{center}
 \includegraphics[scale=1]{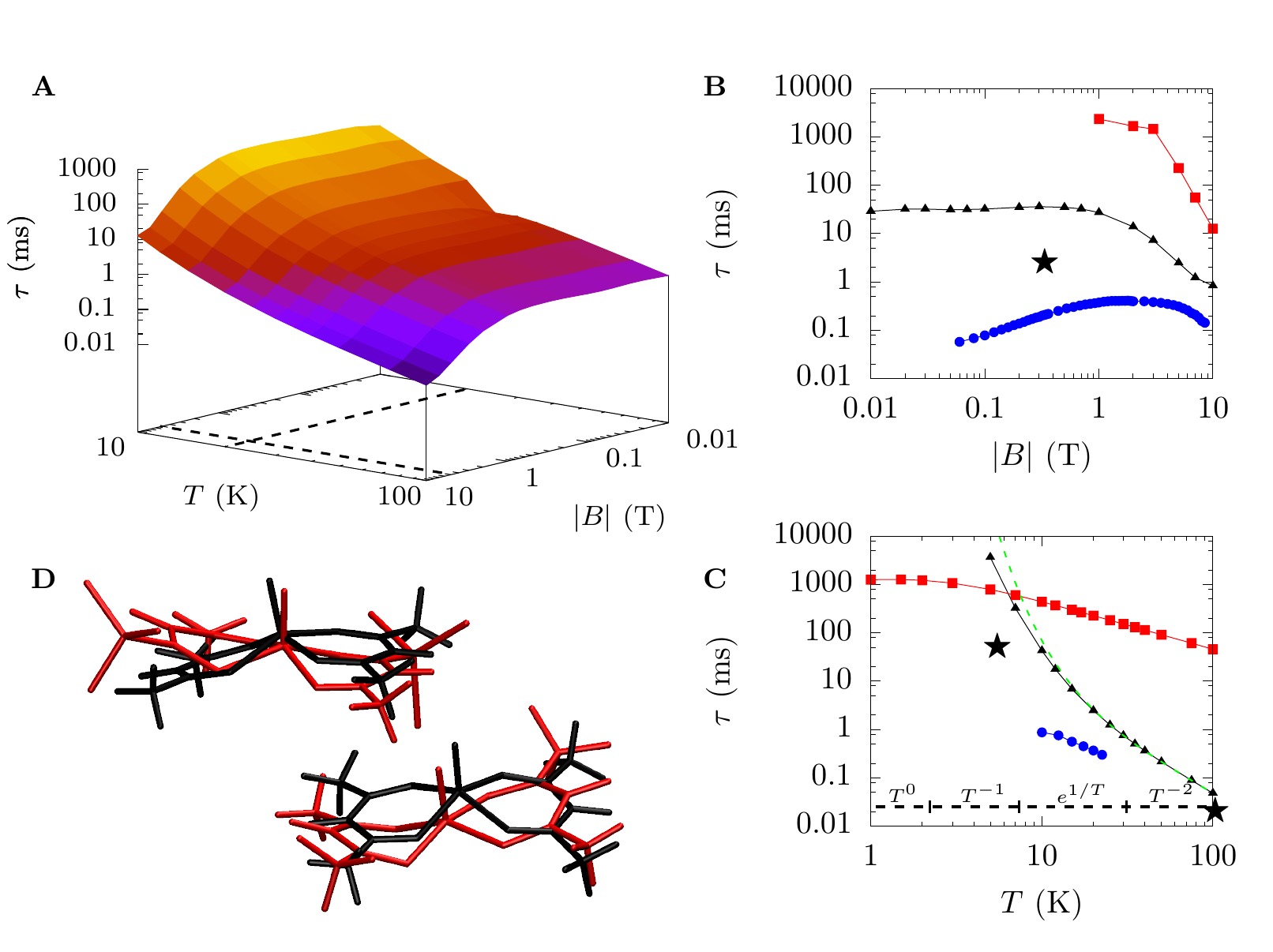}
\end{center}
\caption{\textbf{Spin-phonon relaxation time.} (\textbf{A}) Spin lifetime due to Raman relaxation as a 
function of the external magnetic field, $|B|$, and the temperature, $T$. The dashed black lines on 
the $BT$ plane are the path used for the one-dimensional plots reported in panels (\textbf{B},  \textbf{C}). 
(\textbf{B}) Computed direct (red line and squares), computed Raman (black line and triangles) and AC 
measured (blue line and circles) spin-relaxation times as function of the external magnetic field $|B|$ at 
20~K. The star corresponds to the average $T_{1}$ obtained from EPR measurements of magnetically 
diluted samples, both solid-state and frozen solution~\cite{Atzori2016,Yu2016,Atzori2016a,Atzori2018a}. 
(\textbf{C}) Computed direct (red line and squares), computed Raman (black line and triangles) and AC 
measured (blue line and circles) spin-relaxation times as function of the temperature $T$ at an external 
magnetic field of 5~T. The star corresponds to the average $T_{1}$ obtained from X-Band EPR measurements 
of diluted samples, both solid-state and frozen solution~\cite{Atzori2016,Yu2016,Atzori2016a,Atzori2018a}. 
The dashed green line corresponds to the fit of the second term of Eq.~\ref{local}. The dashed black line and labels help individuating the different $T$ regimes. (\textbf{D}) Cartesian displacements associated with the first and second optical modes at the $\Gamma$-point. The black structure is the VO(acac)$_{2}$ equilibrium geometry as obtained from periodic-DFT optimization.}
\label{scheme3}
\end{figure*}

$G^{1-\mathrm{ph}}$ has been investigated previously~\cite{Lunghi2019b} starting from the force 
constants of the 3$\times$3$\times$3 super-cell in Fig.~\ref{scheme2}A calculated with density functional 
theory (DFT). At the VO(acac)$_{2}$ characteristic frequencies the V spin is in resonance with acoustic 
phonons. These modes have a tiny phonon density of state (DOS), as their $\mathbf{q}$ approaches the 
$\Gamma$-point, but, at the same time, possess a large thermal population due to their low-energy 
($\omega_{\alpha\mathbf{q}}\ll k_\mathrm{B}T$). In order to obtain similar insights into Raman-driven relaxation, 
we use the DFT force constants and analyse $\bar{G}^{2-\mathrm{ph}}=\sum_{\mathbf{q}\mathbf{q'}}\delta(\omega-\omega_{\alpha \mathbf{q}}+\omega_{\beta \mathbf{q'}}) n_{\alpha \mathbf{q}}(n_{\beta \mathbf{q'}}+1)$, where 
the integration is over the Brillouin zone. Fig.~\ref{scheme2}C shows $\bar{G}^{2-\mathrm{ph}}$ as function 
of both $\omega=\omega_{\alpha}-\omega_{\beta}$ and $\omega_{\beta}$. The maximum of this distribution 
shows the range of phonon energies, $\omega_{\beta}$, having a predominant contribution to the two-phonon 
DOS for a given value of $\omega$, and helps individuating the region of the phonons' spectrum responsible 
for the spin relaxation. The maximum contribution to the low-energy $\bar{G}^{2-\mathrm{ph}}$ for 
$\omega$ close to the energy of available spin transitions, is provided by absorption/emission of 
phonons at $\sim40$~cm$^{-1}$ with virtually no effect from phonons above 70~cm$^{-1}$ 
and below 20~cm$^{-1}$. Thus, while in principles the entire vibrational spectrum contributes to 
$\bar{G}^{2-\mathrm{ph}}$, only a portion of the phonon's spectrum has, at the same time, a large 
DOS and thermal population. 

In order to determine the spin-relaxation rate, associated to one- and two-phonon processes, we 
compute the corresponding spin-phonon coupling coefficients with supervised machine learning (ML). 
Our ML method consists in encoding the molecular structure into atomic-environment fingerprints, 
the bispectrum components~\cite{Bartok2013}, and use them together with Ridge regression 
to predict the spin-Hamiltonian coefficients~\cite{Lunghi2019,Lunghi2020} (see Fig.~\ref{scheme2}B). 
Less than 2000 DFT reference data are enough to successfully train a ML model able to predict 
the eigenvalues of the $\mathbf{g_{e}}$ and $\mathbf{A}$ tensors with a statistical error of 
only $\sim2\cdot10^{-5}$ units and $\sim4.2\cdot10^{-2}$~MHz, respectively. 
Once the ML model is trained, it is possible to rapidly scan \textbf{A} and $\mathbf{g_{e}}$ along 
every molecular degrees of freedom and take the first- and second-order numerical derivatives at 
a negligible computational cost. An example of two-dimensional scans of \textbf{A} and $\mathbf{g_{e}}$ 
is reported in Fig.~\ref{scheme2}D. The relatively flat surface of the spin-Hamiltonian parameters 
nicely shows the appropriateness of the weak spin-phonon coupling regime assumed by the Redfield 
theory, a fact never demonstrated before. We stress that the second-order numerical differentiation of 
$\mathbf{A}$ and $\mathbf{g_{e}}$ requires the prediction of the spin-Hamiltonian coefficients for over 
$10^{5}$ molecular geometries. This volume of calculations is hardly sustainable with ab-initio methods, 
but becomes feasible on a large scale with ML.

Fig.~\ref{scheme3}A reports the spin-relaxation time $\tau$ due to two-phonon excitations as a function of 
the external magnetic field and temperature. The hyperfine interaction is responsible for driving Raman 
relaxation at $B<1$~T, regardless of the temperature range, while Zeeman interaction is effective at 
higher fields. Figs.~\ref{scheme3}B and C present a comparison between $\tau$ due to direct and Raman 
mechanisms at constant temperature, $T=20$~K, and constant field, $|B|=5$~T. Raman spin relaxation 
presents a $B^{-2}$ dependence at fields where the Zeeman interaction mediates the relaxation, while it 
becomes field independent at lower fields, where hyperfine interaction dominates. At the relatively high 
temperature of 20~K, Raman relaxation dominates over the entire field window investigated. Nevertheless, 
the direct relaxation due to Zeeman interaction has a $B^{-4}$ field dependence and it is expected to take 
over the Raman mechanism at larger fields. At temperatures below 40~K Raman relaxation becomes 
exponentially slow due to the lack of thermal phonons and the direct relaxation mechanism takes over with 
a $T^{-1}$ dependence. In the ultra-low $T$ regime spin relaxation becomes independent of $T$ and 
proceeds through the spontaneous emission of a single phonon from an excited spin state. Finally, for 
$T>40 K$, $\tau$ reaches a $T^{-2}$ dependence. 

Figures \ref{scheme3}B and \ref{scheme3}C also report the comparison with AC experimental data \cite{Tesi2016}. 
A good agreement is observed for high-$T$ high-field conditions, where the dipolar interactions are less 
effective or absent. Since experiments on magnetically diluted samples are not available for VO(acac)$_{2}$, 
we compare our simulations with the $T_{1}$ obtained from electron paramagnetic resonance (EPR) on 
magnetically diluted samples of other Vanadyl compounds \cite{Atzori2016,Yu2016,Atzori2016a,Atzori2018a}. 
All these molecules show relaxation times of the same order of magnitude across the temperature range 
5-100~K. A representative value for their relaxation time is reported in Fig.~\ref{scheme3}B and Fig.~\ref{scheme3}C. 
Overall we observe a nice agreement between the simulated and experimental behaviour for this class 
of molecular qubits. The overestimation of the experimental results of approximately one order of magnitude 
is consistent with our overestimation of the vibrational frequencies and the underestimations of the hyperfine constants \cite{Lunghi2019b}. We expect that further optimization of our first-principles methods and the inclusion of 
dipolar interactions will be able to further improve the agreement with experiments.

The temperature dependence of the spin-relaxation rate can be modelled with the expression,
\begin{equation}
\tau^{-1}=V_{1-\mathrm{ph}}\frac{e^{\beta\omega_{1-\mathrm{ph}}}}{(e^{\beta\omega_{1-\mathrm{ph}}}-1)}+V_{2-\mathrm{ph}}\frac{e^{\beta\omega_{2-\mathrm{ph}}}} {  (e^{\beta\omega_{2-\mathrm{ph}}}-1)^{2}}\:, 
\label{local}
\end{equation}
where $\beta=(k_\mathrm{B}T)^{-1}$. The second term is suggested by the product 
$\bar{n}_{\alpha\mathbf{q}}(\bar{n}_{\beta\mathbf{q'}}+1)$ present in $G^{2-\mathrm{ph}}$ under the condition 
$\bar{n}_{\alpha\mathbf{q}}\sim\bar{n}_{\beta\mathbf{q'}}$ and can be used to identify the phonons responsible 
for Raman relaxation. A regression of the Raman relaxation time with the second term of Eq.~(\ref{local}), 
using $V_{2-\mathrm{ph}}$ and $\omega_{2-\mathrm{ph}}$ as free parameters, returns 
$\omega_{2-\mathrm{ph}}\sim 44$~cm$^{-1}$ for the dynamics at $|B|=5$T, in nice agreement with the analysis 
done for $\bar{G}^{2-\mathrm{ph}}$. Phonons at such frequency are found across the entire Brillouin zone 
and are reminiscent of the first few $\Gamma$-point modes. The molecular distortions associated to the first 
two optical modes at the $\Gamma$-point are reported in Fig.~\ref{scheme3}D and show a strong contribution 
from Methyl rotations and from twisting of the acac ligands with respect to the Vanadyl unit. 

These findings complement the significant experimental efforts that have been devoted to the study of the relation between vibrations and spin relaxation in recent years \cite{Astner2018,Atzori2018a,Moseley2018,Garlatti2020} and conclusively point to low-energy vibrations and soft intra-molecular modes as key to understand both Direct and Raman spin relaxation in molecular qubits.

The understanding of spin relaxation revealed by our first-principles calculations is rather different from the canonical Raman models, where the relaxation arises from uniform coupling of the spin to Debye-like acoustic modes across the entire Brillouin zone. This is expected to follow a $T^{-9}$ power law \cite{Shrivastava1983a}. Such $T$-dependence is consistently not observed in molecular systems in favour of a $T^{-n}$ power law with $n$ often in the range 2-4 at high-$T$ \cite{Ariciu2019}. Here we reproduce such behaviour and we reinterpret the origin of Raman spin-phonon relaxation as due to the modulation of magnetic interactions in the MHz-GHz energy range by a group of THz intra-molecular vibrations. These results thus support Klemens' interpretation of Raman relaxation as due to local vibrations \cite{Klemens1962}. In addition it offers a new mean to interpret the existing vast spin-relaxation phenomenology in magnetic molecules and possibly improve their properties. 

Taking a look at other solid-state spin qubits, such as organic radicals \cite{Schott2017}, solid-state defects \cite{Astner2018,Lombardi2019} and atoms adsorbed on surfaces \cite{Yang2019a}, we note that they all likely present local vibrations with energy in the THz window able to modulate local magnetic interactions and initiate spin relaxation. Therefore we expect that the analysis offered here for VO(acac)$_{2}$ also teaches important lessons for these systems. Finally, from a methodological point of view, given the recent great advancements of electronic structure methods \cite{Neese2019} and ML for materials science \cite{Butler2018}, the delivery of an accurate and feasible computational approach to the prediction of spin-relaxation time in general classes of materials seems within the reach of the approach proposed here.

\vspace{0.2cm}
\noindent
\textbf{Acknowledgements}

\noindent
This work has been sponsored by AMBER (grant 12/RC/2278\_P2). Computational resources were 
provided by the Trinity Centre for High Performance Computing (TCHPC) and the Irish Centre for 
High-End Computing (ICHEC). We also acknowledge the MOLSPIN COST action CA15128.

\vspace{0.2cm}
\noindent
\textbf{Conflict of interests}\\
The authors declare no competing interests.


\begin{thebibliography}{33}%
\makeatletter
\providecommand \@ifxundefined [1]{%
 \@ifx{#1\undefined}
}%
\providecommand \@ifnum [1]{%
 \ifnum #1\expandafter \@firstoftwo
 \else \expandafter \@secondoftwo
 \fi
}%
\providecommand \@ifx [1]{%
 \ifx #1\expandafter \@firstoftwo
 \else \expandafter \@secondoftwo
 \fi
}%
\providecommand \natexlab [1]{#1}%
\providecommand \enquote  [1]{``#1''}%
\providecommand \bibnamefont  [1]{#1}%
\providecommand \bibfnamefont [1]{#1}%
\providecommand \citenamefont [1]{#1}%
\providecommand \href@noop [0]{\@secondoftwo}%
\providecommand \href [0]{\begingroup \@sanitize@url \@href}%
\providecommand \@href[1]{\@@startlink{#1}\@@href}%
\providecommand \@@href[1]{\endgroup#1\@@endlink}%
\providecommand \@sanitize@url [0]{\catcode `\\12\catcode `\$12\catcode
  `\&12\catcode `\#12\catcode `\^12\catcode `\_12\catcode `\%12\relax}%
\providecommand \@@startlink[1]{}%
\providecommand \@@endlink[0]{}%
\providecommand \url  [0]{\begingroup\@sanitize@url \@url }%
\providecommand \@url [1]{\endgroup\@href {#1}{\urlprefix }}%
\providecommand \urlprefix  [0]{URL }%
\providecommand \Eprint [0]{\href }%
\providecommand \doibase [0]{http://dx.doi.org/}%
\providecommand \selectlanguage [0]{\@gobble}%
\providecommand \bibinfo  [0]{\@secondoftwo}%
\providecommand \bibfield  [0]{\@secondoftwo}%
\providecommand \translation [1]{[#1]}%
\providecommand \BibitemOpen [0]{}%
\providecommand \bibitemStop [0]{}%
\providecommand \bibitemNoStop [0]{.\EOS\space}%
\providecommand \EOS [0]{\spacefactor3000\relax}%
\providecommand \BibitemShut  [1]{\csname bibitem#1\endcsname}%
\let\auto@bib@innerbib\@empty
%</preamble>
\bibitem [{\citenamefont {Hornberger}\ \emph {et~al.}(2004)\citenamefont
  {Hornberger}, \citenamefont {Zeilinger},\ and\ \citenamefont
  {Arndt}}]{Hornberger2004}%
  \BibitemOpen
  \bibfield  {author} {\bibinfo {author} {\bibfnamefont {Klaus}\ \bibnamefont
  {Hornberger}}, \bibinfo {author} {\bibfnamefont {Anton}\ \bibnamefont
  {Zeilinger}}, \ and\ \bibinfo {author} {\bibfnamefont {Markus}\ \bibnamefont
  {Arndt}},\ }\bibfield  {title} {\enquote {\bibinfo {title} {{Decoherence of
  matter waves by thermal emission of radiation}},}\ }\href@noop {} {\bibfield
  {journal} {\bibinfo  {journal} {Nature}\ }\textbf {\bibinfo {volume} {427}},\
  \bibinfo {pages} {711--714} (\bibinfo {year} {2004})}\BibitemShut {NoStop}%
\bibitem [{\citenamefont {Awschalom}\ \emph {et~al.}(2013)\citenamefont
  {Awschalom}, \citenamefont {Basset}, \citenamefont {Dzurak}, \citenamefont
  {Hu},\ and\ \citenamefont {Petta}}]{Awschalom2013}%
  \BibitemOpen
  \bibfield  {author} {\bibinfo {author} {\bibfnamefont {David~D}\ \bibnamefont
  {Awschalom}}, \bibinfo {author} {\bibfnamefont {Lee~C}\ \bibnamefont
  {Basset}}, \bibinfo {author} {\bibfnamefont {Andrew~S}\ \bibnamefont
  {Dzurak}}, \bibinfo {author} {\bibfnamefont {Evelyn~L}\ \bibnamefont {Hu}}, \
  and\ \bibinfo {author} {\bibfnamefont {Jason~R}\ \bibnamefont {Petta}},\
  }\bibfield  {title} {\enquote {\bibinfo {title} {{Quantum Spintronics:
  Engineering and Manipulating Atom-Like Spins in Semiconductors}},}\ }\href
  {\doibase 10.1126/science.1231364} {\bibfield  {journal} {\bibinfo  {journal}
  {Science}\ }\textbf {\bibinfo {volume} {339}},\ \bibinfo {pages} {1174--1179}
  (\bibinfo {year} {2013})}\BibitemShut {NoStop}%
\bibitem [{\citenamefont {Wernsdorfer}\ and\ \citenamefont
  {Ruben}(2019)}]{Wernsdorfer2019}%
  \BibitemOpen
  \bibfield  {author} {\bibinfo {author} {\bibfnamefont {Wolfgang}\
  \bibnamefont {Wernsdorfer}}\ and\ \bibinfo {author} {\bibfnamefont {Mario}\
  \bibnamefont {Ruben}},\ }\bibfield  {title} {\enquote {\bibinfo {title}
  {{Synthetic Hilbert Space Engineering of Molecular Qudits: Isotopologue
  Chemistry}},}\ }\href {\doibase 10.1002/adma.201806687} {\bibfield  {journal}
  {\bibinfo  {journal} {Adv. Mater.}\ }\textbf {\bibinfo {volume} {31}},\
  \bibinfo {pages} {1806687} (\bibinfo {year} {2019})}\BibitemShut {NoStop}%
\bibitem [{\citenamefont {Zadrozny}\ \emph {et~al.}(2015)\citenamefont
  {Zadrozny}, \citenamefont {Niklas}, \citenamefont {Poluektov},\ and\
  \citenamefont {Freedman}}]{Zadrozny2015}%
  \BibitemOpen
  \bibfield  {author} {\bibinfo {author} {\bibfnamefont {Joseph~M.}\
  \bibnamefont {Zadrozny}}, \bibinfo {author} {\bibfnamefont {Jens}\
  \bibnamefont {Niklas}}, \bibinfo {author} {\bibfnamefont {Oleg~G.}\
  \bibnamefont {Poluektov}}, \ and\ \bibinfo {author} {\bibfnamefont
  {Danna~E.}\ \bibnamefont {Freedman}},\ }\bibfield  {title} {\enquote
  {\bibinfo {title} {{Millisecond Coherence Time in a Tunable Molecular
  Electronic Spin Qubit}},}\ }\href {\doibase 10.1021/acscentsci.5b00338}
  {\bibfield  {journal} {\bibinfo  {journal} {ACS Cent. Sci.}\ }\textbf
  {\bibinfo {volume} {1}},\ \bibinfo {pages} {488--492} (\bibinfo {year}
  {2015})}\BibitemShut {NoStop}%
\bibitem [{\citenamefont {Shiddiq}\ \emph {et~al.}(2016)\citenamefont
  {Shiddiq}, \citenamefont {Komijani}, \citenamefont {Duan}, \citenamefont
  {Gaita-Ari{\~{n}}o}, \citenamefont {Coronado},\ and\ \citenamefont
  {Hill}}]{Shiddiq2016}%
  \BibitemOpen
  \bibfield  {author} {\bibinfo {author} {\bibfnamefont {Muhandis}\
  \bibnamefont {Shiddiq}}, \bibinfo {author} {\bibfnamefont {Dorsa}\
  \bibnamefont {Komijani}}, \bibinfo {author} {\bibfnamefont {Yan}\
  \bibnamefont {Duan}}, \bibinfo {author} {\bibfnamefont {Alejandro}\
  \bibnamefont {Gaita-Ari{\~{n}}o}}, \bibinfo {author} {\bibfnamefont
  {Eugenio}\ \bibnamefont {Coronado}}, \ and\ \bibinfo {author} {\bibfnamefont
  {Stephen}\ \bibnamefont {Hill}},\ }\bibfield  {title} {\enquote {\bibinfo
  {title} {{Enhancing coherence in molecular spin qubits via atomic clock
  transitions}},}\ }\href {\doibase 10.1038/nature16984} {\bibfield  {journal}
  {\bibinfo  {journal} {Nature}\ }\textbf {\bibinfo {volume} {531}},\ \bibinfo
  {pages} {348--351} (\bibinfo {year} {2016})}\BibitemShut {NoStop}%
\bibitem [{\citenamefont {Maehrlein}\ \emph {et~al.}(2018)\citenamefont
  {Maehrlein}, \citenamefont {Radu}, \citenamefont {Maldonado}, \citenamefont
  {Paarmann}, \citenamefont {Gensch}, \citenamefont {Kalashnikova},
  \citenamefont {Pisarev}, \citenamefont {Wolf}, \citenamefont {Oppeneer},
  \citenamefont {Barker},\ and\ \citenamefont {Kampfrath}}]{Maehrlein2018}%
  \BibitemOpen
  \bibfield  {author} {\bibinfo {author} {\bibfnamefont {Sebastian~F}\
  \bibnamefont {Maehrlein}}, \bibinfo {author} {\bibfnamefont {Ilie}\
  \bibnamefont {Radu}}, \bibinfo {author} {\bibfnamefont {Pablo}\ \bibnamefont
  {Maldonado}}, \bibinfo {author} {\bibfnamefont {Alexander}\ \bibnamefont
  {Paarmann}}, \bibinfo {author} {\bibfnamefont {Michael}\ \bibnamefont
  {Gensch}}, \bibinfo {author} {\bibfnamefont {Alexandra~M}\ \bibnamefont
  {Kalashnikova}}, \bibinfo {author} {\bibfnamefont {Roman~V}\ \bibnamefont
  {Pisarev}}, \bibinfo {author} {\bibfnamefont {Martin}\ \bibnamefont {Wolf}},
  \bibinfo {author} {\bibfnamefont {Peter~M}\ \bibnamefont {Oppeneer}},
  \bibinfo {author} {\bibfnamefont {Joseph}\ \bibnamefont {Barker}}, \ and\
  \bibinfo {author} {\bibfnamefont {Tobias}\ \bibnamefont {Kampfrath}},\
  }\bibfield  {title} {\enquote {\bibinfo {title} {{Dissecting spin-phonon
  equilibration in ferrimagnetic insulators by ultrafast lattice
  excitation}},}\ }\href@noop {} {\bibfield  {journal} {\bibinfo  {journal}
  {Sci. Adv.}\ }\textbf {\bibinfo {volume} {4}},\ \bibinfo {pages} {eaar5164}
  (\bibinfo {year} {2018})}\BibitemShut {NoStop}%
\bibitem [{\citenamefont {Astner}\ \emph {et~al.}(2018)\citenamefont {Astner},
  \citenamefont {Gugler}, \citenamefont {Angerer}, \citenamefont {Wald},
  \citenamefont {Putz}, \citenamefont {Mauser}, \citenamefont {Trupke},
  \citenamefont {Sumiya}, \citenamefont {Onoda}, \citenamefont {Isoya},
  \citenamefont {Schmiedmayer}, \citenamefont {Mohn},\ and\ \citenamefont
  {Majer}}]{Astner2018}%
  \BibitemOpen
  \bibfield  {author} {\bibinfo {author} {\bibfnamefont {T.}~\bibnamefont
  {Astner}}, \bibinfo {author} {\bibfnamefont {J.}~\bibnamefont {Gugler}},
  \bibinfo {author} {\bibfnamefont {A.}~\bibnamefont {Angerer}}, \bibinfo
  {author} {\bibfnamefont {S.}~\bibnamefont {Wald}}, \bibinfo {author}
  {\bibfnamefont {S.}~\bibnamefont {Putz}}, \bibinfo {author} {\bibfnamefont
  {N.~J.}\ \bibnamefont {Mauser}}, \bibinfo {author} {\bibfnamefont
  {M.}~\bibnamefont {Trupke}}, \bibinfo {author} {\bibfnamefont
  {H.}~\bibnamefont {Sumiya}}, \bibinfo {author} {\bibfnamefont
  {S.}~\bibnamefont {Onoda}}, \bibinfo {author} {\bibfnamefont
  {J.}~\bibnamefont {Isoya}}, \bibinfo {author} {\bibfnamefont
  {J.}~\bibnamefont {Schmiedmayer}}, \bibinfo {author} {\bibfnamefont
  {P.}~\bibnamefont {Mohn}}, \ and\ \bibinfo {author} {\bibfnamefont
  {J.}~\bibnamefont {Majer}},\ }\bibfield  {title} {\enquote {\bibinfo {title}
  {{Solid-state electron spin lifetime limited by phononic vacuum modes}},}\
  }\href@noop {} {\bibfield  {journal} {\bibinfo  {journal} {Nat. Mater.}\
  }\textbf {\bibinfo {volume} {17}},\ \bibinfo {pages} {313--317} (\bibinfo
  {year} {2018})}\BibitemShut {NoStop}%
\bibitem [{\citenamefont {Caravan}(2006)}]{Caravan2006}%
  \BibitemOpen
  \bibfield  {author} {\bibinfo {author} {\bibfnamefont {Peter}\ \bibnamefont
  {Caravan}},\ }\bibfield  {title} {\enquote {\bibinfo {title} {{Strategies for
  increasing the sensitivity of gadolinium based MRI contrast agents}},}\
  }\href@noop {} {\bibfield  {journal} {\bibinfo  {journal} {Chem. Soc. Rev}\
  }\textbf {\bibinfo {volume} {35}},\ \bibinfo {pages} {512--523} (\bibinfo
  {year} {2006})}\BibitemShut {NoStop}%
\bibitem [{\citenamefont {Escalera-Moreno}\ \emph {et~al.}(2017)\citenamefont
  {Escalera-Moreno}, \citenamefont {Suaud}, \citenamefont {Gaita-Ari{\~{n}}o},\
  and\ \citenamefont {Coronado}}]{Escalera-Moreno2017}%
  \BibitemOpen
  \bibfield  {author} {\bibinfo {author} {\bibfnamefont {L.}~\bibnamefont
  {Escalera-Moreno}}, \bibinfo {author} {\bibfnamefont {N.}~\bibnamefont
  {Suaud}}, \bibinfo {author} {\bibfnamefont {A.}~\bibnamefont
  {Gaita-Ari{\~{n}}o}}, \ and\ \bibinfo {author} {\bibfnamefont
  {E.}~\bibnamefont {Coronado}},\ }\bibfield  {title} {\enquote {\bibinfo
  {title} {{Determining Key Local Vibrations in the Relaxation of Molecular
  Spin Qubits and Single-Molecule Magnets}},}\ }\href {\doibase
  10.1021/acs.jpclett.7b00479} {\bibfield  {journal} {\bibinfo  {journal} {J.
  Phys. Chem. Lett.}\ }\textbf {\bibinfo {volume} {8}},\ \bibinfo {pages}
  {1695--1700} (\bibinfo {year} {2017})}\BibitemShut {NoStop}%
\bibitem [{\citenamefont {Goodwin}\ \emph {et~al.}(2017)\citenamefont
  {Goodwin}, \citenamefont {Ortu}, \citenamefont {Reta}, \citenamefont
  {Chilton},\ and\ \citenamefont {Mills}}]{Goodwin2017}%
  \BibitemOpen
  \bibfield  {author} {\bibinfo {author} {\bibfnamefont {Conrad~A.P.}\
  \bibnamefont {Goodwin}}, \bibinfo {author} {\bibfnamefont {Fabrizio}\
  \bibnamefont {Ortu}}, \bibinfo {author} {\bibfnamefont {Daniel}\ \bibnamefont
  {Reta}}, \bibinfo {author} {\bibfnamefont {Nicholas~F.}\ \bibnamefont
  {Chilton}}, \ and\ \bibinfo {author} {\bibfnamefont {David~P.}\ \bibnamefont
  {Mills}},\ }\bibfield  {title} {\enquote {\bibinfo {title} {{Molecular
  magnetic hysteresis at 60 kelvin in dysprosocenium}},}\ }\href {\doibase
  10.1038/nature23447} {\bibfield  {journal} {\bibinfo  {journal} {Nature}\
  }\textbf {\bibinfo {volume} {548}},\ \bibinfo {pages} {439--442} (\bibinfo
  {year} {2017})}\BibitemShut {NoStop}%
\bibitem [{\citenamefont {Lunghi}\ and\ \citenamefont
  {Sanvito}(2019{\natexlab{a}})}]{Lunghi2019b}%
  \BibitemOpen
  \bibfield  {author} {\bibinfo {author} {\bibfnamefont {Alessandro}\
  \bibnamefont {Lunghi}}\ and\ \bibinfo {author} {\bibfnamefont {Stefano}\
  \bibnamefont {Sanvito}},\ }\bibfield  {title} {\enquote {\bibinfo {title}
  {{How do phonons relax molecular spins?}}}\ }\href@noop {} {\bibfield
  {journal} {\bibinfo  {journal} {Sci. Adv.}\ }\textbf {\bibinfo {volume}
  {5}},\ \bibinfo {pages} {eaax7163} (\bibinfo {year}
  {2019}{\natexlab{a}})}\BibitemShut {NoStop}%
\bibitem [{\citenamefont {Lunghi}\ and\ \citenamefont
  {Sanvito}(2019{\natexlab{b}})}]{Lunghi2019}%
  \BibitemOpen
  \bibfield  {author} {\bibinfo {author} {\bibfnamefont {Alessandro}\
  \bibnamefont {Lunghi}}\ and\ \bibinfo {author} {\bibfnamefont {Stefano}\
  \bibnamefont {Sanvito}},\ }\bibfield  {title} {\enquote {\bibinfo {title} {{A
  unified picture of the covalent bond within quantum-accurate force fields :
  From organic molecules to metallic complexes ' reactivity}},}\ }\href@noop {}
  {\bibfield  {journal} {\bibinfo  {journal} {Sci. Adv.}\ }\textbf {\bibinfo
  {volume} {5}},\ \bibinfo {pages} {eaaw2210} (\bibinfo {year}
  {2019}{\natexlab{b}})}\BibitemShut {NoStop}%
\bibitem [{\citenamefont {Lunghi}\ and\ \citenamefont
  {Sanvito}(2020)}]{Lunghi2020}%
  \BibitemOpen
  \bibfield  {author} {\bibinfo {author} {\bibfnamefont {Alessandro}\
  \bibnamefont {Lunghi}}\ and\ \bibinfo {author} {\bibfnamefont {Stefano}\
  \bibnamefont {Sanvito}},\ }\bibfield  {title} {\enquote {\bibinfo {title}
  {{Surfing Multiple Conformation-Property Landscapes via Machine Learning:
  Designing Single-Ion Magnetic Anisotropy}},}\ }\href {\doibase
  10.1021/acs.jpcc.0c01187} {\bibfield  {journal} {\bibinfo  {journal} {J.
  Phys. Chem. C}\ }\textbf {\bibinfo {volume} {124}},\ \bibinfo {pages}
  {5802--5806} (\bibinfo {year} {2020})}\BibitemShut {NoStop}%
\bibitem [{\citenamefont {Tesi}\ \emph {et~al.}(2016)\citenamefont {Tesi},
  \citenamefont {Lunghi}, \citenamefont {Atzori}, \citenamefont {Lucaccini},
  \citenamefont {Sorace}, \citenamefont {Totti},\ and\ \citenamefont
  {Sessoli}}]{Tesi2016}%
  \BibitemOpen
  \bibfield  {author} {\bibinfo {author} {\bibfnamefont {Lorenzo}\ \bibnamefont
  {Tesi}}, \bibinfo {author} {\bibfnamefont {Alessandro}\ \bibnamefont
  {Lunghi}}, \bibinfo {author} {\bibfnamefont {Matteo}\ \bibnamefont {Atzori}},
  \bibinfo {author} {\bibfnamefont {Eva}\ \bibnamefont {Lucaccini}}, \bibinfo
  {author} {\bibfnamefont {Lorenzo}\ \bibnamefont {Sorace}}, \bibinfo {author}
  {\bibfnamefont {Federico}\ \bibnamefont {Totti}}, \ and\ \bibinfo {author}
  {\bibfnamefont {Roberta}\ \bibnamefont {Sessoli}},\ }\bibfield  {title}
  {\enquote {\bibinfo {title} {{Giant spin-phonon bottleneck effects in
  evaporable vanadyl-based molecules with long spin coherence}},}\ }\href
  {\doibase 10.1039/C6DT02559E} {\bibfield  {journal} {\bibinfo  {journal}
  {Dalt. Trans.}\ }\textbf {\bibinfo {volume} {45}},\ \bibinfo {pages}
  {16635--16645} (\bibinfo {year} {2016})}\BibitemShut {NoStop}%
\bibitem [{\citenamefont {Childress}\ \emph {et~al.}(2006)\citenamefont
  {Childress}, \citenamefont {{Gurudev Dutt}}, \citenamefont {Taylor},
  \citenamefont {Zibrov}, \citenamefont {Jelezko}, \citenamefont {Wrachtrup},
  \citenamefont {Hemmer},\ and\ \citenamefont {Lukin}}]{Childress2006}%
  \BibitemOpen
  \bibfield  {author} {\bibinfo {author} {\bibfnamefont {L}~\bibnamefont
  {Childress}}, \bibinfo {author} {\bibfnamefont {M~V}\ \bibnamefont {{Gurudev
  Dutt}}}, \bibinfo {author} {\bibfnamefont {J~M}\ \bibnamefont {Taylor}},
  \bibinfo {author} {\bibfnamefont {A~S}\ \bibnamefont {Zibrov}}, \bibinfo
  {author} {\bibfnamefont {F}~\bibnamefont {Jelezko}}, \bibinfo {author}
  {\bibfnamefont {J}~\bibnamefont {Wrachtrup}}, \bibinfo {author}
  {\bibfnamefont {P~R}\ \bibnamefont {Hemmer}}, \ and\ \bibinfo {author}
  {\bibfnamefont {M~D}\ \bibnamefont {Lukin}},\ }\bibfield  {title} {\enquote
  {\bibinfo {title} {{Coherent Dynamics of Coupled Electron and Nuclear Spin
  Qubits in Diamond}},}\ }\href@noop {} {\bibfield  {journal} {\bibinfo
  {journal} {Science}\ }\textbf {\bibinfo {volume} {314}},\ \bibinfo {pages}
  {281--286} (\bibinfo {year} {2006})}\BibitemShut {NoStop}%
\bibitem [{\citenamefont {Muhonen}\ \emph {et~al.}(2014)\citenamefont
  {Muhonen}, \citenamefont {Dehollain}, \citenamefont {Laucht}, \citenamefont
  {Hudson}, \citenamefont {Kalra}, \citenamefont {Sekiguchi}, \citenamefont
  {Itoh}, \citenamefont {Jamieson}, \citenamefont {McCallum}, \citenamefont
  {Dzurak},\ and\ \citenamefont {Morello}}]{Muhonen2014}%
  \BibitemOpen
  \bibfield  {author} {\bibinfo {author} {\bibfnamefont {Juha~T}\ \bibnamefont
  {Muhonen}}, \bibinfo {author} {\bibfnamefont {Juan~P}\ \bibnamefont
  {Dehollain}}, \bibinfo {author} {\bibfnamefont {Arne}\ \bibnamefont
  {Laucht}}, \bibinfo {author} {\bibfnamefont {Fay~E}\ \bibnamefont {Hudson}},
  \bibinfo {author} {\bibfnamefont {Rachpon}\ \bibnamefont {Kalra}}, \bibinfo
  {author} {\bibfnamefont {Takeharu}\ \bibnamefont {Sekiguchi}}, \bibinfo
  {author} {\bibfnamefont {Kohei~M}\ \bibnamefont {Itoh}}, \bibinfo {author}
  {\bibfnamefont {David~N}\ \bibnamefont {Jamieson}}, \bibinfo {author}
  {\bibfnamefont {Jeffrey~C}\ \bibnamefont {McCallum}}, \bibinfo {author}
  {\bibfnamefont {Andrew~S}\ \bibnamefont {Dzurak}}, \ and\ \bibinfo {author}
  {\bibfnamefont {Andrea}\ \bibnamefont {Morello}},\ }\bibfield  {title}
  {\enquote {\bibinfo {title} {{Storing quantum information for 30 seconds in a
  nanoelectronic device}},}\ }\href@noop {} {\bibfield  {journal} {\bibinfo
  {journal} {Nat. Nanotechnol.}\ }\textbf {\bibinfo {volume} {9}},\ \bibinfo
  {pages} {986} (\bibinfo {year} {2014})}\BibitemShut {NoStop}%
\bibitem [{\citenamefont {Godfrin}\ \emph {et~al.}(2017)\citenamefont
  {Godfrin}, \citenamefont {Ferhat}, \citenamefont {Ballou}, \citenamefont
  {Klyatskaya}, \citenamefont {Ruben}, \citenamefont {Wernsdorfer},\ and\
  \citenamefont {Balestro}}]{Godfrin2017a}%
  \BibitemOpen
  \bibfield  {author} {\bibinfo {author} {\bibfnamefont {C}~\bibnamefont
  {Godfrin}}, \bibinfo {author} {\bibfnamefont {A}~\bibnamefont {Ferhat}},
  \bibinfo {author} {\bibfnamefont {R}~\bibnamefont {Ballou}}, \bibinfo
  {author} {\bibfnamefont {S}~\bibnamefont {Klyatskaya}}, \bibinfo {author}
  {\bibfnamefont {M}~\bibnamefont {Ruben}}, \bibinfo {author} {\bibfnamefont
  {W}~\bibnamefont {Wernsdorfer}}, \ and\ \bibinfo {author} {\bibfnamefont
  {F}~\bibnamefont {Balestro}},\ }\bibfield  {title} {\enquote {\bibinfo
  {title} {{Operating Quantum States in Single Magnetic Molecules:
  Implementation of Grover's Quantum Algorithm}},}\ }\href@noop {} {\bibfield
  {journal} {\bibinfo  {journal} {Phys. Rev. Lett.}\ }\textbf {\bibinfo
  {volume} {119}},\ \bibinfo {pages} {187702} (\bibinfo {year}
  {2017})}\BibitemShut {NoStop}%
\bibitem [{\citenamefont {Redfield}(1965)}]{Redfield1965}%
  \BibitemOpen
  \bibfield  {author} {\bibinfo {author} {\bibfnamefont {A~G}\ \bibnamefont
  {Redfield}},\ }\href {\doibase 10.1016/B978-1-4832-3114-3.50007-6} {\emph
  {\bibinfo {title} {Adv. Magn. Opt. Reson.}}},\ Vol.~\bibinfo {volume} {1}\
  (\bibinfo  {publisher} {Academic Press Inc.},\ \bibinfo {year} {1965})\ pp.\
  \bibinfo {pages} {1--32}\BibitemShut {NoStop}%
\bibitem [{\citenamefont {Atzori}\ \emph
  {et~al.}(2016{\natexlab{a}})\citenamefont {Atzori}, \citenamefont {Tesi},
  \citenamefont {Morra}, \citenamefont {Chiesa}, \citenamefont {Sorace},\ and\
  \citenamefont {Sessoli}}]{Atzori2016}%
  \BibitemOpen
  \bibfield  {author} {\bibinfo {author} {\bibfnamefont {Matteo}\ \bibnamefont
  {Atzori}}, \bibinfo {author} {\bibfnamefont {Lorenzo}\ \bibnamefont {Tesi}},
  \bibinfo {author} {\bibfnamefont {Elena}\ \bibnamefont {Morra}}, \bibinfo
  {author} {\bibfnamefont {Mario}\ \bibnamefont {Chiesa}}, \bibinfo {author}
  {\bibfnamefont {Lorenzo}\ \bibnamefont {Sorace}}, \ and\ \bibinfo {author}
  {\bibfnamefont {Roberta}\ \bibnamefont {Sessoli}},\ }\bibfield  {title}
  {\enquote {\bibinfo {title} {{Room-Temperature Quantum Coherence and Rabi
  Oscillations in Vanadyl Phthalocyanine: Toward Multifunctional Molecular Spin
  Qubits}},}\ }\href {\doibase 10.1021/jacs.5b13408} {\bibfield  {journal}
  {\bibinfo  {journal} {J. Am. Chem. Soc.}\ }\textbf {\bibinfo {volume}
  {138}},\ \bibinfo {pages} {2154--2157} (\bibinfo {year}
  {2016}{\natexlab{a}})}\BibitemShut {NoStop}%
\bibitem [{\citenamefont {Yu}\ \emph {et~al.}(2016)\citenamefont {Yu},
  \citenamefont {Graham}, \citenamefont {Zadrozny}, \citenamefont {Niklas},
  \citenamefont {Krzyaniak}, \citenamefont {Wasielewski}, \citenamefont
  {Poluektov},\ and\ \citenamefont {Freedman}}]{Yu2016}%
  \BibitemOpen
  \bibfield  {author} {\bibinfo {author} {\bibfnamefont {Chung~Jui}\
  \bibnamefont {Yu}}, \bibinfo {author} {\bibfnamefont {Michael~J.}\
  \bibnamefont {Graham}}, \bibinfo {author} {\bibfnamefont {Joseph~M.}\
  \bibnamefont {Zadrozny}}, \bibinfo {author} {\bibfnamefont {Jens}\
  \bibnamefont {Niklas}}, \bibinfo {author} {\bibfnamefont {Matthew~D.}\
  \bibnamefont {Krzyaniak}}, \bibinfo {author} {\bibfnamefont {Michael~R.}\
  \bibnamefont {Wasielewski}}, \bibinfo {author} {\bibfnamefont {Oleg~G.}\
  \bibnamefont {Poluektov}}, \ and\ \bibinfo {author} {\bibfnamefont
  {Danna~E.}\ \bibnamefont {Freedman}},\ }\bibfield  {title} {\enquote
  {\bibinfo {title} {{Long Coherence Times in Nuclear Spin-Free Vanadyl
  Qubits}},}\ }\href {\doibase 10.1021/jacs.6b08467} {\bibfield  {journal}
  {\bibinfo  {journal} {J. Am. Chem. Soc.}\ }\textbf {\bibinfo {volume}
  {138}},\ \bibinfo {pages} {14678--14685} (\bibinfo {year}
  {2016})}\BibitemShut {NoStop}%
\bibitem [{\citenamefont {Atzori}\ \emph
  {et~al.}(2016{\natexlab{b}})\citenamefont {Atzori}, \citenamefont {Morra},
  \citenamefont {Tesi}, \citenamefont {Albino}, \citenamefont {Chiesa},
  \citenamefont {Sorace},\ and\ \citenamefont {Sessoli}}]{Atzori2016a}%
  \BibitemOpen
  \bibfield  {author} {\bibinfo {author} {\bibfnamefont {Matteo}\ \bibnamefont
  {Atzori}}, \bibinfo {author} {\bibfnamefont {Elena}\ \bibnamefont {Morra}},
  \bibinfo {author} {\bibfnamefont {Lorenzo}\ \bibnamefont {Tesi}}, \bibinfo
  {author} {\bibfnamefont {Andrea}\ \bibnamefont {Albino}}, \bibinfo {author}
  {\bibfnamefont {Mario}\ \bibnamefont {Chiesa}}, \bibinfo {author}
  {\bibfnamefont {Lorenzo}\ \bibnamefont {Sorace}}, \ and\ \bibinfo {author}
  {\bibfnamefont {Roberta}\ \bibnamefont {Sessoli}},\ }\bibfield  {title}
  {\enquote {\bibinfo {title} {{Quantum Coherence Times Enhancement in
  Vanadium(IV)-based Potential Molecular Qubits: The Key Role of the Vanadyl
  Moiety}},}\ }\href {\doibase 10.1021/jacs.6b05574} {\bibfield  {journal}
  {\bibinfo  {journal} {J. Am. Chem. Soc.}\ }\textbf {\bibinfo {volume}
  {138}},\ \bibinfo {pages} {11234--11244} (\bibinfo {year}
  {2016}{\natexlab{b}})}\BibitemShut {NoStop}%
\bibitem [{\citenamefont {Atzori}\ \emph {et~al.}(2018)\citenamefont {Atzori},
  \citenamefont {Benci}, \citenamefont {Morra}, \citenamefont {Tesi},
  \citenamefont {Chiesa}, \citenamefont {Torre}, \citenamefont {Sorace},\ and\
  \citenamefont {Sessoli}}]{Atzori2018a}%
  \BibitemOpen
  \bibfield  {author} {\bibinfo {author} {\bibfnamefont {Matteo}\ \bibnamefont
  {Atzori}}, \bibinfo {author} {\bibfnamefont {Stefano}\ \bibnamefont {Benci}},
  \bibinfo {author} {\bibfnamefont {Elena}\ \bibnamefont {Morra}}, \bibinfo
  {author} {\bibfnamefont {Lorenzo}\ \bibnamefont {Tesi}}, \bibinfo {author}
  {\bibfnamefont {Mario}\ \bibnamefont {Chiesa}}, \bibinfo {author}
  {\bibfnamefont {Renato}\ \bibnamefont {Torre}}, \bibinfo {author}
  {\bibfnamefont {Lorenzo}\ \bibnamefont {Sorace}}, \ and\ \bibinfo {author}
  {\bibfnamefont {Roberta}\ \bibnamefont {Sessoli}},\ }\bibfield  {title}
  {\enquote {\bibinfo {title} {{Structural Effects on the Spin Dynamics of
  Potential Molecular Qubits}},}\ }\href {\doibase
  10.1021/acs.inorgchem.7b02616} {\bibfield  {journal} {\bibinfo  {journal}
  {Inorg. Chem.}\ }\textbf {\bibinfo {volume} {57}},\ \bibinfo {pages}
  {731--740} (\bibinfo {year} {2018})}\BibitemShut {NoStop}%
\bibitem [{\citenamefont {Bartok}\ \emph {et~al.}(2013)\citenamefont {Bartok},
  \citenamefont {Kondor},\ and\ \citenamefont {Csanyi}}]{Bartok2013}%
  \BibitemOpen
  \bibfield  {author} {\bibinfo {author} {\bibfnamefont {Albert~P.}\
  \bibnamefont {Bartok}}, \bibinfo {author} {\bibfnamefont {Risi}\ \bibnamefont
  {Kondor}}, \ and\ \bibinfo {author} {\bibfnamefont {Gabor}\ \bibnamefont
  {Csanyi}},\ }\bibfield  {title} {\enquote {\bibinfo {title} {{On representing
  chemical environments}},}\ }\href {\doibase 10.1103/PhysRevB.87.184115}
  {\bibfield  {journal} {\bibinfo  {journal} {Phys. Rev. B}\ }\textbf {\bibinfo
  {volume} {87}},\ \bibinfo {pages} {184115} (\bibinfo {year}
  {2013})}\BibitemShut {NoStop}%
\bibitem [{\citenamefont {Moseley}\ \emph {et~al.}(2018)\citenamefont
  {Moseley}, \citenamefont {Stavretis}, \citenamefont {Thirunavukkuarasu},
  \citenamefont {Ozerov}, \citenamefont {Cheng}, \citenamefont {Daemen},
  \citenamefont {Ludwig}, \citenamefont {Lu}, \citenamefont {Smirnov},
  \citenamefont {Brown}, \citenamefont {Pandey}, \citenamefont
  {Ramirez-Cuesta}, \citenamefont {Lamb}, \citenamefont {Atanasov},
  \citenamefont {Bill}, \citenamefont {Neese},\ and\ \citenamefont
  {Xue}}]{Moseley2018}%
  \BibitemOpen
  \bibfield  {author} {\bibinfo {author} {\bibfnamefont {Duncan~H.}\
  \bibnamefont {Moseley}}, \bibinfo {author} {\bibfnamefont {Shelby~E.}\
  \bibnamefont {Stavretis}}, \bibinfo {author} {\bibfnamefont {Komalavalli}\
  \bibnamefont {Thirunavukkuarasu}}, \bibinfo {author} {\bibfnamefont
  {Mykhaylo}\ \bibnamefont {Ozerov}}, \bibinfo {author} {\bibfnamefont
  {Yongqiang}\ \bibnamefont {Cheng}}, \bibinfo {author} {\bibfnamefont
  {Luke~L.}\ \bibnamefont {Daemen}}, \bibinfo {author} {\bibfnamefont
  {Jonathan}\ \bibnamefont {Ludwig}}, \bibinfo {author} {\bibfnamefont
  {Zhengguang}\ \bibnamefont {Lu}}, \bibinfo {author} {\bibfnamefont {Dmitry}\
  \bibnamefont {Smirnov}}, \bibinfo {author} {\bibfnamefont {Craig~M.}\
  \bibnamefont {Brown}}, \bibinfo {author} {\bibfnamefont {Anup}\ \bibnamefont
  {Pandey}}, \bibinfo {author} {\bibfnamefont {A.~J.}\ \bibnamefont
  {Ramirez-Cuesta}}, \bibinfo {author} {\bibfnamefont {Adam~C.}\ \bibnamefont
  {Lamb}}, \bibinfo {author} {\bibfnamefont {Mihail}\ \bibnamefont {Atanasov}},
  \bibinfo {author} {\bibfnamefont {Eckhard}\ \bibnamefont {Bill}}, \bibinfo
  {author} {\bibfnamefont {Frank}\ \bibnamefont {Neese}}, \ and\ \bibinfo
  {author} {\bibfnamefont {Zi~Ling}\ \bibnamefont {Xue}},\ }\bibfield  {title}
  {\enquote {\bibinfo {title} {{Spin-phonon couplings in transition metal
  complexes with slow magnetic relaxation}},}\ }\href {\doibase
  10.1038/s41467-018-04896-0} {\bibfield  {journal} {\bibinfo  {journal} {Nat.
  Commun.}\ }\textbf {\bibinfo {volume} {9}},\ \bibinfo {pages} {2572}
  (\bibinfo {year} {2018})}\BibitemShut {NoStop}%
\bibitem [{\citenamefont {Garlatti}\ \emph {et~al.}(2020)\citenamefont
  {Garlatti}, \citenamefont {Tesi}, \citenamefont {Lunghi}, \citenamefont
  {Atzori}, \citenamefont {Voneshen}, \citenamefont {Santini}, \citenamefont
  {Sanvito}, \citenamefont {Guidi}, \citenamefont {Sessoli},\ and\
  \citenamefont {Carretta}}]{Garlatti2020}%
  \BibitemOpen
  \bibfield  {author} {\bibinfo {author} {\bibfnamefont {E}~\bibnamefont
  {Garlatti}}, \bibinfo {author} {\bibfnamefont {L}~\bibnamefont {Tesi}},
  \bibinfo {author} {\bibfnamefont {A}~\bibnamefont {Lunghi}}, \bibinfo
  {author} {\bibfnamefont {M}~\bibnamefont {Atzori}}, \bibinfo {author}
  {\bibfnamefont {D~J}\ \bibnamefont {Voneshen}}, \bibinfo {author}
  {\bibfnamefont {P}~\bibnamefont {Santini}}, \bibinfo {author} {\bibfnamefont
  {S}~\bibnamefont {Sanvito}}, \bibinfo {author} {\bibfnamefont
  {T}~\bibnamefont {Guidi}}, \bibinfo {author} {\bibfnamefont {R}~\bibnamefont
  {Sessoli}}, \ and\ \bibinfo {author} {\bibfnamefont {S}~\bibnamefont
  {Carretta}},\ }\bibfield  {title} {\enquote {\bibinfo {title} {{Unveiling
  phonons in a molecular qubit with four-dimensional inelastic neutron
  scattering and density functional theory}},}\ }\href@noop {} {\bibfield
  {journal} {\bibinfo  {journal} {Nat. Commun.}\ ,\ \bibinfo {pages}
  {DOI:0.1038/s41467--020--15475--7}} (\bibinfo {year} {2020})}\BibitemShut
  {NoStop}%
\bibitem [{\citenamefont {Shrivastava}(1983)}]{Shrivastava1983a}%
  \BibitemOpen
  \bibfield  {author} {\bibinfo {author} {\bibfnamefont {K~N}\ \bibnamefont
  {Shrivastava}},\ }\bibfield  {title} {\enquote {\bibinfo {title} {{Theory of
  Spin-Lattice Relaxation}},}\ }\href@noop {} {\bibfield  {journal} {\bibinfo
  {journal} {Phys. Status Solidi}\ }\textbf {\bibinfo {volume} {117}},\
  \bibinfo {pages} {437--458} (\bibinfo {year} {1983})}\BibitemShut {NoStop}%
\bibitem [{\citenamefont {Ariciu}\ \emph {et~al.}(2019)\citenamefont {Ariciu},
  \citenamefont {Woen}, \citenamefont {Huh}, \citenamefont {Nodaraki},
  \citenamefont {Kostopoulos}, \citenamefont {Goodwin}, \citenamefont
  {Chilton}, \citenamefont {Mcinnes}, \citenamefont {Winpenny}, \citenamefont
  {Evans},\ and\ \citenamefont {Tuna}}]{Ariciu2019}%
  \BibitemOpen
  \bibfield  {author} {\bibinfo {author} {\bibfnamefont {Ana-Maria}\
  \bibnamefont {Ariciu}}, \bibinfo {author} {\bibfnamefont {David~H}\
  \bibnamefont {Woen}}, \bibinfo {author} {\bibfnamefont {Daniel~N}\
  \bibnamefont {Huh}}, \bibinfo {author} {\bibfnamefont {Lydia~E}\ \bibnamefont
  {Nodaraki}}, \bibinfo {author} {\bibfnamefont {Andreas~K}\ \bibnamefont
  {Kostopoulos}}, \bibinfo {author} {\bibfnamefont {Conrad A~P}\ \bibnamefont
  {Goodwin}}, \bibinfo {author} {\bibfnamefont {Nicholas~F}\ \bibnamefont
  {Chilton}}, \bibinfo {author} {\bibfnamefont {Eric J~L}\ \bibnamefont
  {Mcinnes}}, \bibinfo {author} {\bibfnamefont {Richard E~P}\ \bibnamefont
  {Winpenny}}, \bibinfo {author} {\bibfnamefont {William~J}\ \bibnamefont
  {Evans}}, \ and\ \bibinfo {author} {\bibfnamefont {Floriana}\ \bibnamefont
  {Tuna}},\ }\bibfield  {title} {\enquote {\bibinfo {title} {{Engineering
  electronic structure to prolong relaxation times in molecular qubits by
  minimising orbital angular momentum}},}\ }\href {\doibase
  10.1038/s41467-019-11309-3} {\bibfield  {journal} {\bibinfo  {journal} {Nat.
  Commun.}\ }\textbf {\bibinfo {volume} {10}},\ \bibinfo {pages} {3330}
  (\bibinfo {year} {2019})}\BibitemShut {NoStop}%
\bibitem [{\citenamefont {Klemens}(1962)}]{Klemens1962}%
  \BibitemOpen
  \bibfield  {author} {\bibinfo {author} {\bibfnamefont {P.~G.}\ \bibnamefont
  {Klemens}},\ }\bibfield  {title} {\enquote {\bibinfo {title} {{Localized
  modes and spin-lattice interactions}},}\ }\href {\doibase
  10.1103/PhysRev.125.1795} {\bibfield  {journal} {\bibinfo  {journal} {Phys.
  Rev.}\ }\textbf {\bibinfo {volume} {125}},\ \bibinfo {pages} {1795--1798}
  (\bibinfo {year} {1962})}\BibitemShut {NoStop}%
\bibitem [{\citenamefont {Schott}\ \emph {et~al.}(2017)\citenamefont {Schott},
  \citenamefont {McNellis}, \citenamefont {Nielsen}, \citenamefont {Chen},
  \citenamefont {Watanabe}, \citenamefont {Tanaka}, \citenamefont {McCulloch},
  \citenamefont {Takimiya}, \citenamefont {Sinova},\ and\ \citenamefont
  {Sirringhaus}}]{Schott2017}%
  \BibitemOpen
  \bibfield  {author} {\bibinfo {author} {\bibfnamefont {Sam}\ \bibnamefont
  {Schott}}, \bibinfo {author} {\bibfnamefont {Erik~R}\ \bibnamefont
  {McNellis}}, \bibinfo {author} {\bibfnamefont {Christian~B}\ \bibnamefont
  {Nielsen}}, \bibinfo {author} {\bibfnamefont {Hung-Yang}\ \bibnamefont
  {Chen}}, \bibinfo {author} {\bibfnamefont {Shun}\ \bibnamefont {Watanabe}},
  \bibinfo {author} {\bibfnamefont {Hisaaki}\ \bibnamefont {Tanaka}}, \bibinfo
  {author} {\bibfnamefont {Iain}\ \bibnamefont {McCulloch}}, \bibinfo {author}
  {\bibfnamefont {Kazuo}\ \bibnamefont {Takimiya}}, \bibinfo {author}
  {\bibfnamefont {Jairo}\ \bibnamefont {Sinova}}, \ and\ \bibinfo {author}
  {\bibfnamefont {Henning}\ \bibnamefont {Sirringhaus}},\ }\bibfield  {title}
  {\enquote {\bibinfo {title} {{Tuning the effective spin-orbit coupling in
  molecular semiconductors}},}\ }\href {\doibase 10.1038/ncomms15200}
  {\bibfield  {journal} {\bibinfo  {journal} {Nat. Commun.}\ }\textbf {\bibinfo
  {volume} {8}},\ \bibinfo {pages} {15200} (\bibinfo {year}
  {2017})}\BibitemShut {NoStop}%
\bibitem [{\citenamefont {Lombardi}\ \emph {et~al.}(2019)\citenamefont
  {Lombardi}, \citenamefont {Lodi}, \citenamefont {Ma}, \citenamefont {Liu},
  \citenamefont {Slota}, \citenamefont {Narita}, \citenamefont {Myers},
  \citenamefont {M{\"{u}}llen}, \citenamefont {Feng},\ and\ \citenamefont
  {Bogani}}]{Lombardi2019}%
  \BibitemOpen
  \bibfield  {author} {\bibinfo {author} {\bibfnamefont {Federico}\
  \bibnamefont {Lombardi}}, \bibinfo {author} {\bibfnamefont {Alessandro}\
  \bibnamefont {Lodi}}, \bibinfo {author} {\bibfnamefont {Ji}~\bibnamefont
  {Ma}}, \bibinfo {author} {\bibfnamefont {Junzhi}\ \bibnamefont {Liu}},
  \bibinfo {author} {\bibfnamefont {Michael}\ \bibnamefont {Slota}}, \bibinfo
  {author} {\bibfnamefont {Akimitsu}\ \bibnamefont {Narita}}, \bibinfo {author}
  {\bibfnamefont {William~K}\ \bibnamefont {Myers}}, \bibinfo {author}
  {\bibfnamefont {Klaus}\ \bibnamefont {M{\"{u}}llen}}, \bibinfo {author}
  {\bibfnamefont {Xinliang}\ \bibnamefont {Feng}}, \ and\ \bibinfo {author}
  {\bibfnamefont {Lapo}\ \bibnamefont {Bogani}},\ }\bibfield  {title} {\enquote
  {\bibinfo {title} {{Quantum units from the topological engineering of
  molecular graphenoids}},}\ }\href@noop {} {\bibfield  {journal} {\bibinfo
  {journal} {Science}\ }\textbf {\bibinfo {volume} {366}},\ \bibinfo {pages}
  {1107--1110} (\bibinfo {year} {2019})}\BibitemShut {NoStop}%
\bibitem [{\citenamefont {Yang}\ \emph {et~al.}(2019)\citenamefont {Yang},
  \citenamefont {Paul}, \citenamefont {Phark}, \citenamefont {Willke},
  \citenamefont {Bae}, \citenamefont {Choi}, \citenamefont {Esat},
  \citenamefont {Ardavan}, \citenamefont {Heinrich},\ and\ \citenamefont
  {Lutz}}]{Yang2019a}%
  \BibitemOpen
  \bibfield  {author} {\bibinfo {author} {\bibfnamefont {Kai}\ \bibnamefont
  {Yang}}, \bibinfo {author} {\bibfnamefont {William}\ \bibnamefont {Paul}},
  \bibinfo {author} {\bibfnamefont {Soo-hyon}\ \bibnamefont {Phark}}, \bibinfo
  {author} {\bibfnamefont {Philip}\ \bibnamefont {Willke}}, \bibinfo {author}
  {\bibfnamefont {Yujeong}\ \bibnamefont {Bae}}, \bibinfo {author}
  {\bibfnamefont {Taeyoung}\ \bibnamefont {Choi}}, \bibinfo {author}
  {\bibfnamefont {Taner}\ \bibnamefont {Esat}}, \bibinfo {author}
  {\bibfnamefont {Arzhang}\ \bibnamefont {Ardavan}}, \bibinfo {author}
  {\bibfnamefont {Andreas~J}\ \bibnamefont {Heinrich}}, \ and\ \bibinfo
  {author} {\bibfnamefont {Christopher~P}\ \bibnamefont {Lutz}},\ }\bibfield
  {title} {\enquote {\bibinfo {title} {{Coherent spin manipulation of
  individual atoms on a surface}},}\ }\href@noop {} {\bibfield  {journal}
  {\bibinfo  {journal} {Science}\ }\textbf {\bibinfo {volume} {366}},\ \bibinfo
  {pages} {509--512} (\bibinfo {year} {2019})}\BibitemShut {NoStop}%
\bibitem [{\citenamefont {Neese}\ \emph {et~al.}(2019)\citenamefont {Neese},
  \citenamefont {Atanasov}, \citenamefont {Bistoni}, \citenamefont {Manganas},\
  and\ \citenamefont {Ye}}]{Neese2019}%
  \BibitemOpen
  \bibfield  {author} {\bibinfo {author} {\bibfnamefont {Frank}\ \bibnamefont
  {Neese}}, \bibinfo {author} {\bibfnamefont {Mihail}\ \bibnamefont
  {Atanasov}}, \bibinfo {author} {\bibfnamefont {Giovanni}\ \bibnamefont
  {Bistoni}}, \bibinfo {author} {\bibfnamefont {Dimitrios}\ \bibnamefont
  {Manganas}}, \ and\ \bibinfo {author} {\bibfnamefont {Shengfa}\ \bibnamefont
  {Ye}},\ }\bibfield  {title} {\enquote {\bibinfo {title} {{Chemistry and
  Quantum Mechanics in 2019 – Give us Insight and Numbers}},}\ }\href
  {\doibase 10.1021/jacs.8b13313} {\bibfield  {journal} {\bibinfo  {journal}
  {J. Am. Chem. Soc.}\ }\textbf {\bibinfo {volume} {141}},\ \bibinfo {pages}
  {2814--2824} (\bibinfo {year} {2019})}\BibitemShut {NoStop}%
\bibitem [{\citenamefont {Butler}\ \emph {et~al.}(2018)\citenamefont {Butler},
  \citenamefont {Davies}, \citenamefont {Cartwright}, \citenamefont {Isayev},\
  and\ \citenamefont {Walsh}}]{Butler2018}%
  \BibitemOpen
  \bibfield  {author} {\bibinfo {author} {\bibfnamefont {Keith~T.}\
  \bibnamefont {Butler}}, \bibinfo {author} {\bibfnamefont {Daniel~W.}\
  \bibnamefont {Davies}}, \bibinfo {author} {\bibfnamefont {Hugh}\ \bibnamefont
  {Cartwright}}, \bibinfo {author} {\bibfnamefont {Olexandr}\ \bibnamefont
  {Isayev}}, \ and\ \bibinfo {author} {\bibfnamefont {Aron}\ \bibnamefont
  {Walsh}},\ }\bibfield  {title} {\enquote {\bibinfo {title} {{Machine learning
  for molecular and materials science}},}\ }\href {\doibase
  10.1038/s41586-018-0337-2} {\bibfield  {journal} {\bibinfo  {journal}
  {Nature}\ }\textbf {\bibinfo {volume} {559}},\ \bibinfo {pages} {547--555}
  (\bibinfo {year} {2018})}\BibitemShut {NoStop}%
\end{thebibliography}
\end{document}